\documentclass[aps,prb,preprint,amssymb,superscriptaddress]{revtex4-1}
\usepackage{float}
\usepackage{xcolor}
\usepackage{graphicx}
\usepackage{dcolumn}
\usepackage{bm}
\usepackage{hyperref}
\usepackage{color}
\usepackage{amsmath}
\usepackage{amsfonts}

\begin{document}

\title{Dirac semimetal phase in the hexagonal LiZnBi}

\author{Wendong Cao}
\affiliation{Department of Physics and State Key Laboratory of Low-Dimensional Quantum Physics, Tsinghua University, Beijing 100084, China}
\affiliation{Collaborative Innovation Center of Quantum Matter, Tsinghua University, Beijing 100084, China}

\author{Peizhe Tang}
\affiliation{Department of Physics, McCullough Building, Stanford University, Stanford, California 94305-4045, USA}

\author{Yong Xu}
\affiliation{Department of Physics and State Key Laboratory of Low-Dimensional Quantum Physics, Tsinghua University, Beijing 100084, China}
\affiliation{Collaborative Innovation Center of Quantum Matter, Tsinghua University, Beijing 100084, China}
\affiliation{RIKEN Center for Emergent Matter Science (CEMS), Wako, Saitama 351-0198, Japan}

\author{Jian Wu}
\affiliation{Department of Physics and State Key Laboratory of Low-Dimensional Quantum Physics, Tsinghua University, Beijing 100084, China}
\affiliation{Collaborative Innovation Center of Quantum Matter, Tsinghua University, Beijing 100084, China}

\author{Bing-Lin Gu}
\affiliation{Collaborative Innovation Center of Quantum Matter, Tsinghua University, Beijing 100084, China}
\affiliation{Institute for Advanced Study, Tsinghua University, Beijing 100084, China}

\author{Wenhui Duan}
\email{dwh@phys.tsinghua.edu.cn}
\affiliation{Department of Physics and State Key Laboratory of Low-Dimensional Quantum Physics, Tsinghua University, Beijing 100084, China}
\affiliation{Collaborative Innovation Center of Quantum Matter, Tsinghua University, Beijing 100084, China}
\affiliation{Institute for Advanced Study, Tsinghua University, Beijing 100084, China}

\begin{abstract}
Based on first-principles calculations, we find that LiZnBi, a metallic hexagonal $ABC$ compound, can be driven into a Dirac semimetal with a pair of Dirac points by strain. The nontrivial topological nature of the strained LiZnBi is directly demonstrated by calculating its $\mathbb{Z}_2$ index and the surface states, where the Fermi arcs are clearly observed. The low-energy states as well as topological properties are shown to be sensitive to the strain configurations. The finding of Dirac semimetal phase in LiZnBi may intrigue further researches on the topological properties of hexagonal $ABC$ materials and promote new practical applications.
\end{abstract}
\maketitle
\section{Introduction}
Ferroelectrics have been studied for decades for their unique properties and technical potentials \cite{Scott954}.  Ferroelectric hysteresis, one of the features of ferroelectrics, can be used in information storage, similar to ferromagnetic hysteresis. Several years ago, hexagonal $ABC$ semiconductors were proposed as a new family of materials exhibiting proper ferroelectricity \cite{PhysRevLett.109.167602}. This new family of ferroelectric materials takes a different lattice structure from that of the well-studied $AB$O$_3$ perovskite oxides, which greatly enhances the lattice compatibility of ferroelectrics with other functional materials. Moreover, it is found that some materials of this family are hyperferroelectrics \cite{PhysRevLett.112.127601}. Different from other known proper ferroelectrics, the polarization in hyperferroelectrics persists regardless of depolarization fields, which are common in surfaces or interfaces.

When spin-orbit coupling effect is considered, more interesting phenomena emerge out in this family of compounds. Topological insulators have attracted much interest due to their thrilling properties such as dissipationless metallic surface states and potential techniqual applications in spintronics and quantum computations\cite{RevModPhys.82.3045,RevModPhys.83.1057}. Recent theoretical calculations including spin-orbit coupling effect indicate that KMgBi is a three dimensional (3D) strong topological insulator \cite{arxiv1601.03643}. Recently, as another kind of exotic quantum matter with nontrivial topology, Dirac semimetal, whose valence bands and conduction bands only touch at isolated points (Dirac points), also becomes a hot topic because of its own novel features including Fermi arcs \cite{Xu294} and large linear magnetoresistance \cite{Liang280}, as well as its close relationship with other topologically nontrivial phases (such as Weyl semimetal). Na$_3$Bi \cite{Liu864} and Cd$_3$As$_2$ \cite{Liu677} are two well-studied 3D Dirac semimetal systems. As for the search of more realistic materials, a natural question to ask is whether Dirac semimetal phase can also be found in hexagonal $ABC$ compounds.

In this work, based on first-principles calculations, we show that LiZnBi is a 3D Dirac semimetal under proper strain configurations. By analyzing the components of low-energy electronic states, we find that there exists band inversion at the $\Gamma$ point in strained LiZnBi. We further identify the nontrivial topological nature of the system by both the topological index and nontrivial surface states. Finally, we study the response of low-energy states to different strain configurations and provide the phase diagram of Dirac semimetal in strained LiZnBi.

\section{Methods}
The first-principles calculations are carried out by using density functional theory (DFT) with the projector augmented wave method \cite{PhysRevB.50.17953,PhysRevB.59.1758}, as implemented in the Vienna \textit{ab initio} simulation package \cite{PhysRevB.54.11169}. Plane wave basis set with a kinetic energy cutoff of $\mathrm{300~eV}$ are used. The atomic structure is fully relaxed with $10\times 10\times 6$ Monkhorst-Pack $k$ points and the local-density approximation (LDA) \cite{PhysRevLett.45.566,PhysRevB.23.5048} for the exchange-correlation interactions between electrons, until the residual forces are less than $1\times 10^{-3}~\mathrm{eV/\AA}$. Then the electronic structure is calculated with a modified Becke-Johnson exchange potential \cite{mBJ} where the spin-orbit coupling is included. The maximally localized Wannier functions are obtained by the program Wannier90 \cite{Mostofi2008685}. The Green's function method \cite{0305-4608-15-4-009} is used to calculate the surface electronic spectrum and surface states.

\begin{figure}
\centering
\includegraphics[width=\linewidth]{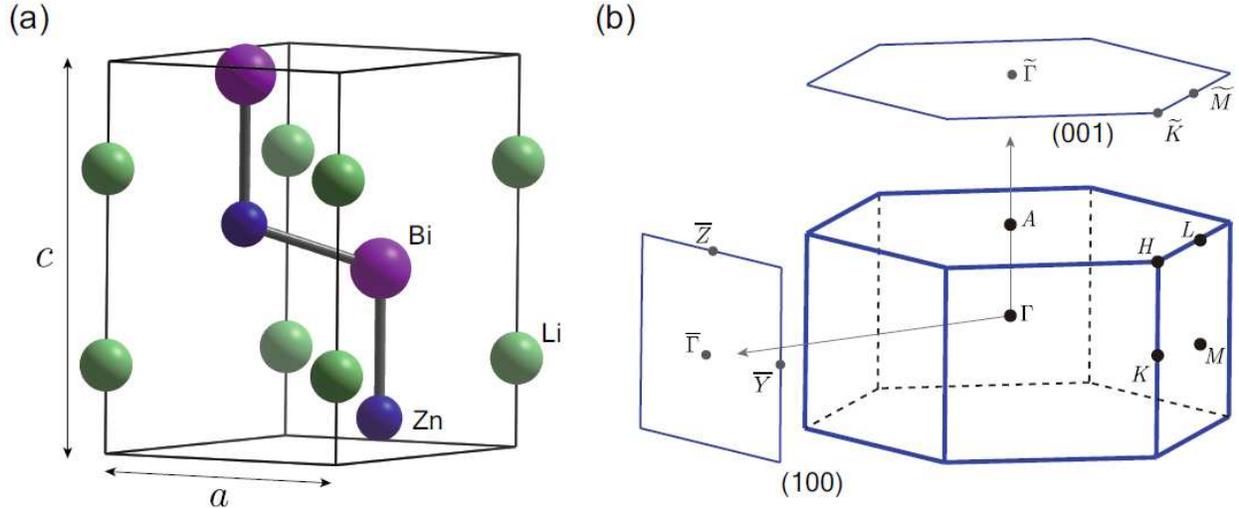}
\caption{(Color online). (a) Lattice structure of LiZnBi with the space group $P6_3mc$, where $a$ and $c$ are two lattice constants. Zn and Bi atoms at Wyckoff positions 2$b$ form a wurtzite structure, while Li atoms are located at 2$a$. (b) The first Brillouin zone and its projections on (100) and (001) surfaces. High-symmetry points are labelled.}
\label{fig:fig1}
\end{figure}
\section{Results}

The lattice structure of LiZnBi, shown in Fig. \ref{fig:fig1} (a), belongs to the space group $P6_3mc$. Li$^+$ ions occupy the  interstitial sites of the wurtzite lattice of ZnBi$^{-}$\cite{PSSA:PSSA200881223}. The calculated lattice constants of the stable structure are $a_0=4.47${\AA}~ and $c_0=7.28$\AA, close to those experimental values ($a_0=4.58${\AA}~ and $c_0=7.38${\AA}) reported in the Inorganic Crystal Structural Database \cite{Belsky:an0615}. The first Brillouin zones of the bulk, (001) surface and (100) surface are shown in Fig. \ref{fig:fig1} (b).

\begin{figure}
\centering
\includegraphics[width=0.8\linewidth]{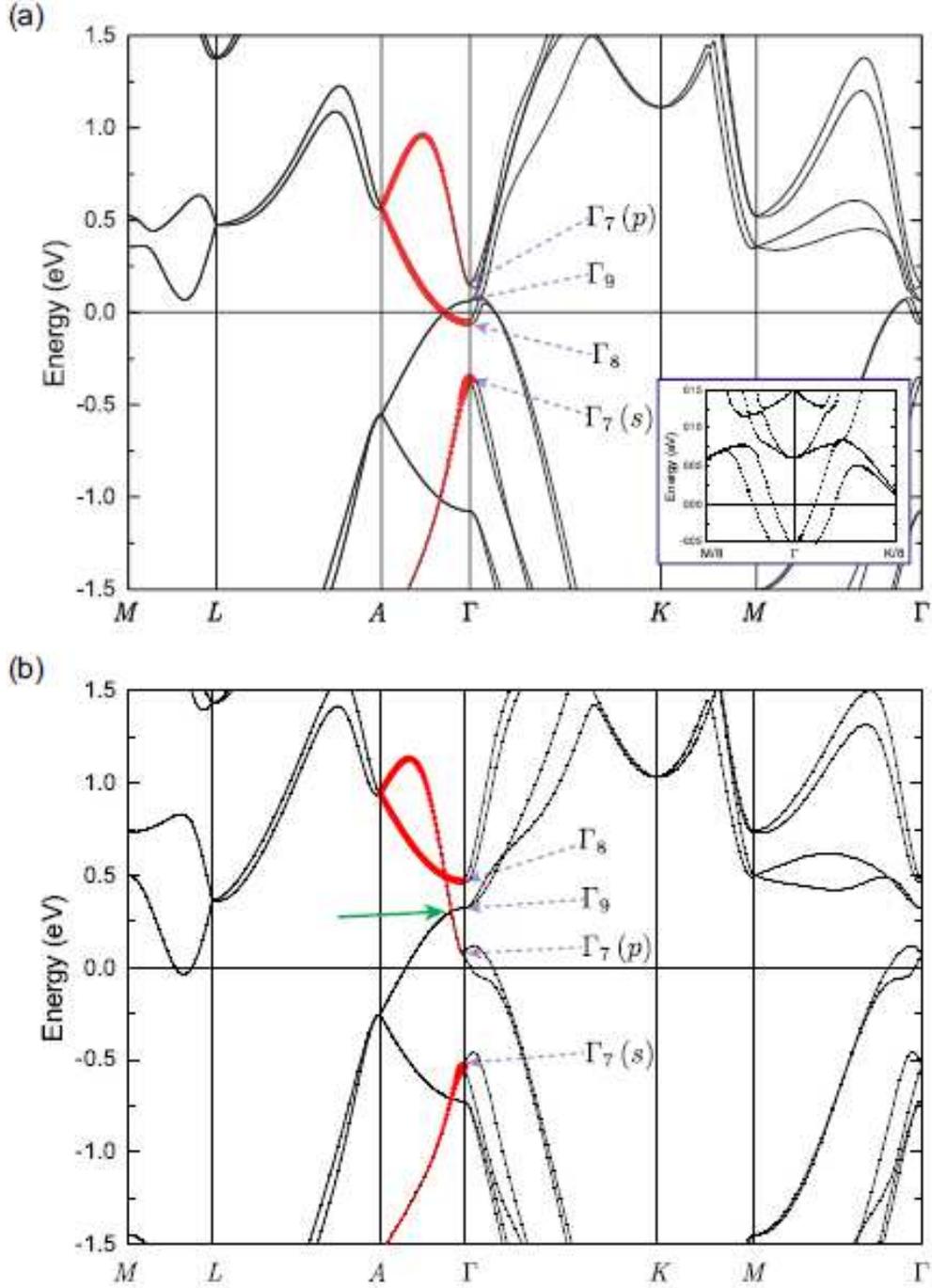}
\caption{(Color online). Band structure along high-symmetry lines of LiZnBi in its unstrained structure (a) and strained structure (b). For (b), $a=0.98a_0$ and $c$ is tuned to keep the volume of the unit cell. In both cases, spin-orbit coupling is included and the Fermi energy is set to zero. The size of red dots along the $A$-$\Gamma$ is proportional to the relative ratio of $s$ orbitals from Zn and Bi atoms. At the $\Gamma$ point, four low-energy states are labelled by their representations. The inset in (a) shows the band structure near the $\Gamma$ point. In (b), the 3D Dirac point is pointed out by the green arrow.}
\label{fig:fig2}
\end{figure}

Figure \ref{fig:fig2} shows the calculated band structure of LiZnBi in its stable (unstrained) and strained structures with spin-orbit coupling considered. In both cases, besides the low conduction bands along the $M$-$L$ line, the low-energy states are mainly located along the $A$-$\Gamma$ line and around the $\Gamma$ point. Notice that even though inversion symmetry is broken in the lattice structure, all the bands along the $A$-$\Gamma$ are doubly degenerate. The double degeneracy is protected by the combination of $C_{6}$ rotation symmetry and $\sigma$ mirror symmetry. The rotation axis is along (001) direction and the reflection plane includes the rotation axis. The key reason is that these two symmetry operators doesn't commute with each other along the $A$-$\Gamma$ when the spin freedom is considered. For the states at the $\Gamma$ point labelled by their representations, our calculations indicate that the $\Gamma_7\,(s)$ and $\Gamma_8$ states are dominated by $s$ orbitals. In contrast, the $\Gamma_9$ and $\Gamma_7\,(p)$ states are mainly contributed by $p_x\pm ip_y$ orbitals of Bi atoms and $p_z$ orbitals of Bi atoms, respectively. For the four corresponding bands along the $A$-$\Gamma$ line, we find that the main components remain the same for $\Gamma_9$ and $\Gamma_8$. For the rest two bands, the varying sizes of red dots imply that the relative ratios of $s$ orbitals from Zn and Bi atoms change as $k$ moves along the $A$-$\Gamma$. The two bands belonging to the same representation $\Gamma_7$ strongly couple with each other. 

For the unstrained structure, the band order at the $\Gamma$ point is $\Gamma_7\,(p)>\Gamma_9>\Gamma_8>\Gamma_7\,(s)$. In conventional semiconductor LiZnSb, $\Gamma_7\,(p),\,\Gamma_9$ are in valence bands and $\Gamma_8,\,\Gamma_7\,(s)$ are in conduction bands \cite{doi:10.1021/ja062526a}. Therefore, compared to LiZnSb, LiZnBi has two sets of band inversion: both $\Gamma_7\,(s)$ and $\Gamma_8$ fall below $\Gamma_7\,(p)$ and $\Gamma_9$. The substitution of Sb atoms by Bi atoms increases the energy of $p$ orbitals from the nitrogen family, pushing up $\Gamma_7\,(p)$ and $\Gamma_9$. The $\Gamma_7\,(p)$ ($J_z=\pm\frac{1}{2}$) state is above $\Gamma_9$ ($J_z=\pm\frac{3}{2}$) [see Fig. \ref{fig:fig2} (a)] due to the $p$-$d$ hybridization induced negative spin-orbit coupling \cite{PhysRevB.90.245308}. Herein $J_z$ denotes the total angular momentum in the $z$ [or (001)] direction. The strong coupling between $\Gamma_7\,(s)$ and $\Gamma_7\,(p)$ pushes their corresponding bands away from the Fermi level along the $A$-$\Gamma$. And the crossing (Dirac point) between the rest two bands is at the Fermi level. Unfortunately, there is no global gap at $\Gamma-K-M$ plane [see the inset of Fig. \ref{fig:fig2} (a)] and neither $\mathbb{Z}_2$ index nor mirror Chern number can be well-defined.

On the other hand, the relative positions of the four energy levels are very sensitive to strain. If the lattice constant $c$ is enlarged and $a$ is reduced, the band order changes as shown in Fig. \ref{fig:fig2} (b) with the new lattice constants presented in the caption. Compared to those in Fig. \ref{fig:fig2} (a), the relative positions of $\Gamma_7\,(p)$ and $\Gamma_8$ are exchanged. As the $p$-$d$ hybridization here mainly involves $p_z$ orbital from Bi atoms, the increased lattice constant along the $z$ direction weakens this effect. So $\Gamma_7\,(p)$ falls below $\Gamma_9$. At the same time, $\Gamma_8$ is pushed above $\Gamma_9$. It is reasonable because for Zn-Bi interactions, the bonding feature along $z$ direction is found in $\Gamma_8$, similar to the case of LiZnSb \cite{doi:10.1021/ja062526a}. Accompanied by the change of band order near the $\Gamma$ point, a global gap at the $\Gamma$-$K$-$M$ plane emerges. The band inversion between $\Gamma_7\,(s)$ and $\Gamma_9$ is found to be topologically nontrivial from the calculated $\mathbb{Z}_2$ index at the $\Gamma$-$K$-$M$ plane \cite{PhysRevB.83.235401} (shown in Appendix A). The crossing at ($0$,$0$,$k_z^D$), pointed by the green arrow in Fig. \ref{fig:fig2}(b), is protected by $C_{6v}$ symmetry and the band dispersions around the crossing point are linear in all momentum directions, leading to a 3D Dirac semimetal. Time-reversal symmetry ensures that the same crossing (Dirac point) also occurs at ($0$,$0$,-$k_z^D$). Therefore, LiZnBi in the strained structure is a Dirac semimetal with a pair of Dirac points near the Fermi level.

\begin{figure}
\centering
\includegraphics[width=\linewidth]{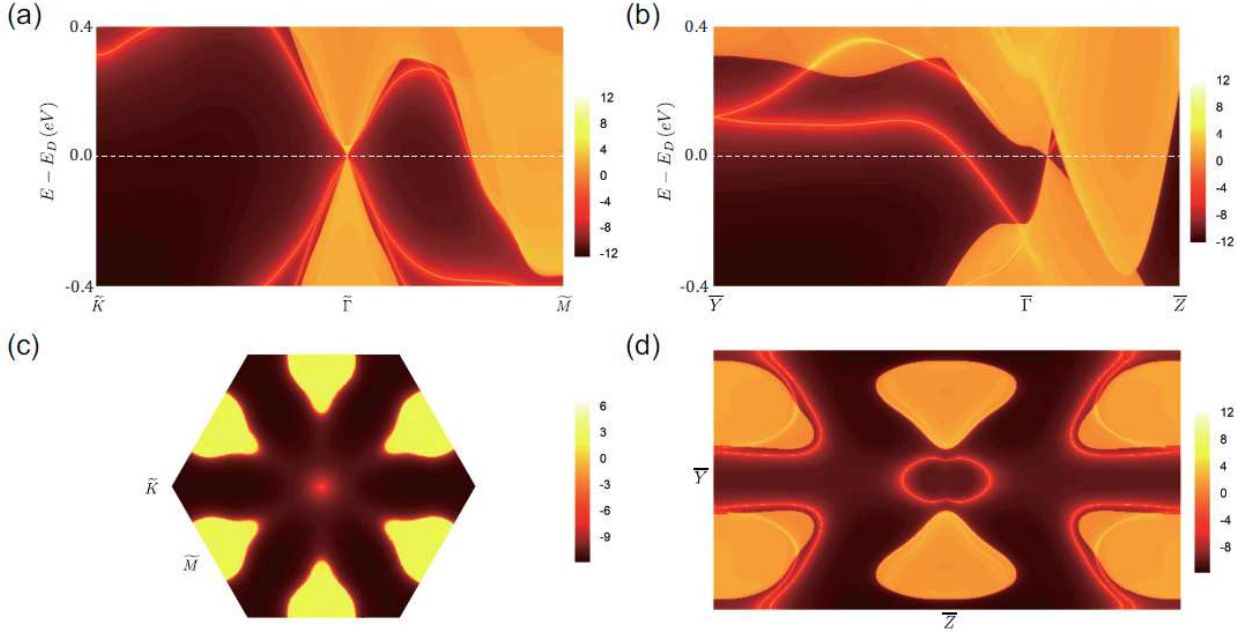}
\caption{(Color online). LDOS for the (001) and (100) surface states of the strained LiZnBi. (a) and (c) are for (001), (b) and (d) are for (100).  In (a) and (b), LDOS along high-symmetry lines are presented where the energy level of the bulk Dirac points, $E_D$, is denoted by dashed white lines. The constant-energy contours at $E_D$ are shown in (c) and (d).}
\label{fig:fig3}
\end{figure}
To further confirm its nontrivial topological nature, we calculate the surface states on (001) and (100) surfaces. Figure \ref{fig:fig3} (a) shows surface states on (001) surface along $\widetilde{K}$-$\widetilde{\Gamma}$-$\widetilde{M}$. Near the $\widetilde{\Gamma}$ point, the linear Dirac cone is from the band dispersion in the $x$-$y$ plane of the 3D Dirac cones. The two bulk Dirac points are projected onto the $\widetilde{\Gamma}$ point in the constant energy contour (see Fig. \ref{fig:fig3} (c)). For the (100) surface, the band dispersion along the $z$ direction of the 3D Dirac cones is projected onto $\overline{\Gamma}$-$\overline{Z}$ and the nontrivial surface states along the $\overline{Y}$-$\overline{\Gamma}$ is consistent with the nonzero $\mathbb{Z}_2$ index at the $\Gamma$-$K$-$M$ plane (see Fig. \ref{fig:fig3} (b)). But as $k$ moves from $\overline{\Gamma}$ to $\overline{Z}$, the band inversion decreases and finally disappears at the projected Dirac points. Correspondingly, the nontrivial surface states exist only within two 3D Dirac points and merge into bulk states beyond that region.  As shown in Fig. \ref{fig:fig3} (d), both the Fermi arcs contributed by the surface states connect the two projected Dirac points and vanish elsewhere, manifesting themselves as a signature of nontrivial topology.

\begin{figure}
\centering
\includegraphics[width=\textwidth]{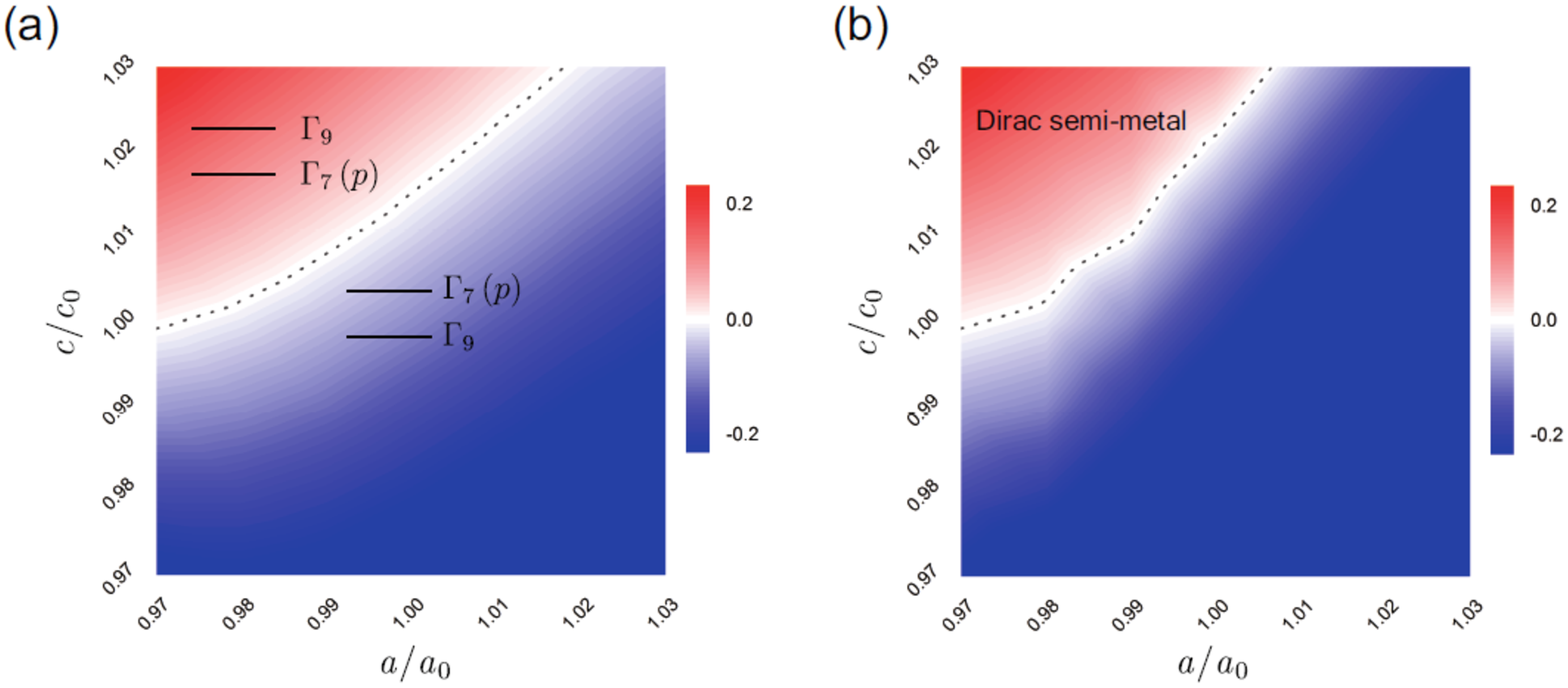}
\caption{(Color online). Low-energy levels at the $\Gamma$ point in LiZnBi with different lattice constants. (a) The effective spin-orbit coupling (i.e. $\Delta_{\text{SOC}}= E[\Gamma_9]-E[\Gamma_7\,(p)]$). The black dotted line denotes the strain conditions, under which the two energy levels are degenerate. (b) The Dirac semimetal phase is in the region where min$\{E[\Gamma_8],E[\Gamma_9]\}-$max$\{E[\Gamma_7\,(p)],E[\Gamma_7\,(s)]\}$ is positive. The phase boundary is shown by the black dotted line. The atomic coordinates are relaxed before electronic structure calculations in all cases.}
\label{fig:fig4}
\end{figure}
Inspired by the change of band order in Fig. \ref{fig:fig2} (b), we further explore the phase boundary of Dirac semimetal by studying the low-energy states under different strain configurations. As shown in Fig. \ref{fig:fig4} (a), the effective spin-orbit coupling strength $\Delta_{\text{SOC}}= E[\Gamma_9]-E[\Gamma_7\,(p)]$ is inclined to turn positive as $a$ decreases or $c$ increases, which may give rise to the appearance of Dirac semimetal phase. Either the smaller in-plane lattice constant $a$ or the larger out-of-plane lattice constant $c$ can weaken the tetrahetral crystal field splitting of $d$ orbitals of Zn and lower the energy levels of $t_{2g}$ orbitals, which plays an important role in the negative effective spin-orbit coupling \cite{PhysRevB.90.245308}. Moreover, $\Gamma_7\,(p)$, mainly consisting of the anti-bonding state of $p_z$ orbitals from Bi atoms and the shallow $d$ orbitals from Zn atoms, gets even lower as the increasing $c$ reduces the bond strength. This trend is consistent with the results shown in Fig. \ref{fig:fig2} where the strained LiZnBi has been proved to be in Dirac semimetal phase. However, since the existence of other low-energy states can affect the occupation of bands, the band order $E[\Gamma_9]>E[\Gamma_7\,(p)]$ itself doesn't ensure the Dirac semimetal phase. In Fig. \ref{fig:fig4} (b), we present the phase boundary of Dirac semimetal (indicated by a black dotted line), which corresponds to zero points of $\min\{E[\Gamma_8],E[\Gamma_9]\}-\max\{E[\Gamma_7\,(p)],E[\Gamma_7\,(s)]\}$. It is reasonable because we find that $\Gamma_8$ is always higher than $\Gamma_7\,(s)$ near the gap closing point ($E[\Gamma_9]-E[\Gamma_7\,(p)]\approx 0$). In fact, various phases can be realized in LiZnBi under different strain configurations. For example, according to our calculations at $a=0.97a_0$ and $c=0.98c_0$ (shown in Appendix B), the strained LiZnBi becomes a conventional semiconductor, suggesting that a strain can induce topological phase transition in LiZnBi.

\section{Conclusions}
In summary, our first-principles calculations indicate that LiZnBi is a 3D Dirac semimetal when both in-plane compressive strain and out-of-plane tenssile strain are applied. We find that there exists band inversion in LiZnBi in both the unstrained and strained structures. But only when a proper strain is applied, the band order changes at the $\Gamma$ point and the band inversion can give rise to nontrivial topology, which is confirmed by the nonzero $\mathbb{Z}_2$ index. Surface states of the strained LiZnBi are also calculated and Fermi arcs are observed on the (100) surface, connecting the two projected Dirac points. Furthermore, we show the phase boundary of Dirac semimetal phase under different strain configurations. Importantly, the critical strain of driving LiZnBi into Dirac semimetal is small and may be realized in experiments by epitaxial growth on proper substrates. For instance, LiZnSb in wurtzite structure can be a good choice for its similar lattice constants and chemical components \cite{Belsky:an0615}.  Our results enrich the topological phases that hexagonal $ABC$ compounds can host and consequently broaden their application potentials.

\begin{acknowledgements}
This work was supported by the Ministry of Science and Technology of China (Grant Nos. 2011CB606405 and 2011CB921901), the National Natural Science Foundation of China (Grant. 11334006). 
\end{acknowledgements}

\appendix
\renewcommand{\theequation}{A.\arabic{equation}}
  \renewcommand{\thetable}{A.\arabic{table}}
\setcounter{equation}{0}  
\setcounter{table}{0}
\renewcommand{\thefigure}{A.\arabic{figure}}
\setcounter{figure}{0}
\section{$\mathbb{Z}_2$ index of the strained LiZnBi in Dirac semimetal phase}
\begin{figure}
\centering
\includegraphics[width=0.8\linewidth]{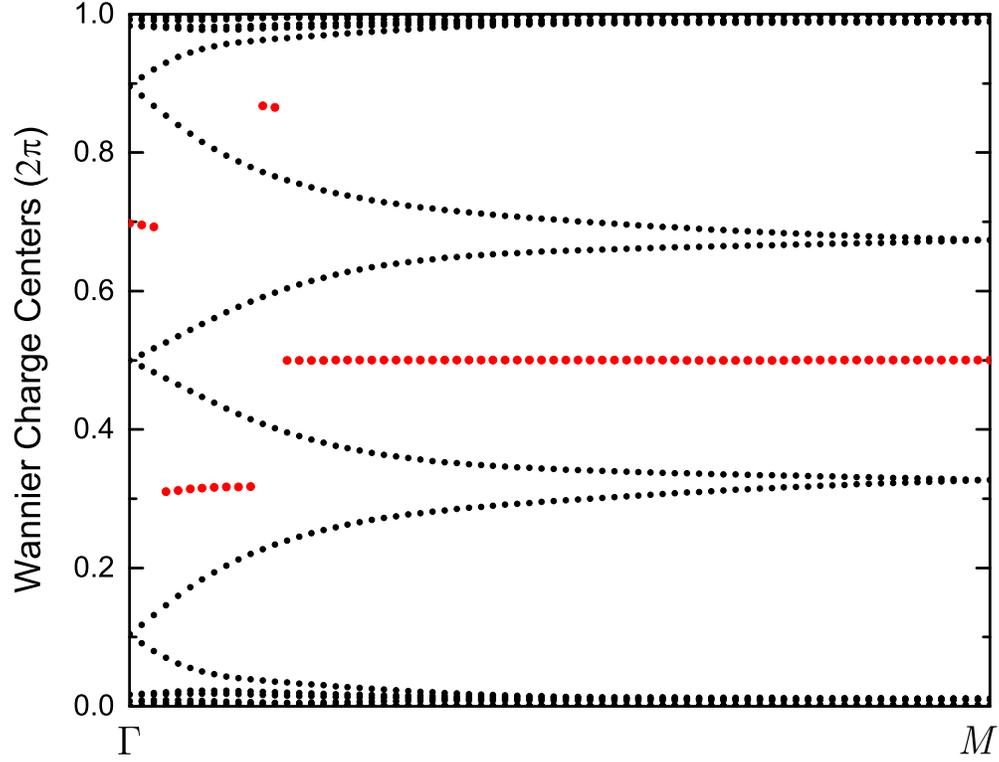}
\caption{(Color online). WCCs along the $\Gamma$-$M$ at the $\Gamma$-$M$-$K$ plane are presented by black dots. The red dots correspond to the center of the largest WCC gap at each $k$ point between $\Gamma$ and $M$.}
\label{fig:appfig1}
\end{figure}

Besides the discussion of band inversion shown in Fig. \ref{fig:fig2} (b), the evaluation of the $\mathbb{Z}_2$ index can directly prove the nontrivial topology in the strained LiZnBi. Since the inversion symmetry is absent, its $\mathbb{Z}_2$ index at $\Gamma$-$M$-$K$ plane is determined by calculating the Wannier Charge Centers (WCCs). The results are shown in Fig. \ref{fig:appfig1}. The total number of WCC bands which the gap center jumps over is seven. It is odd, confirming the nonzero $\mathbb{Z}_2$ index in the strained LiZnBi.

\renewcommand{\thefigure}{B.\arabic{figure}}
\setcounter{figure}{0}
\section{Topological phase transition in strained LiZnBi}
\begin{figure}
\centering
\includegraphics[width=\linewidth]{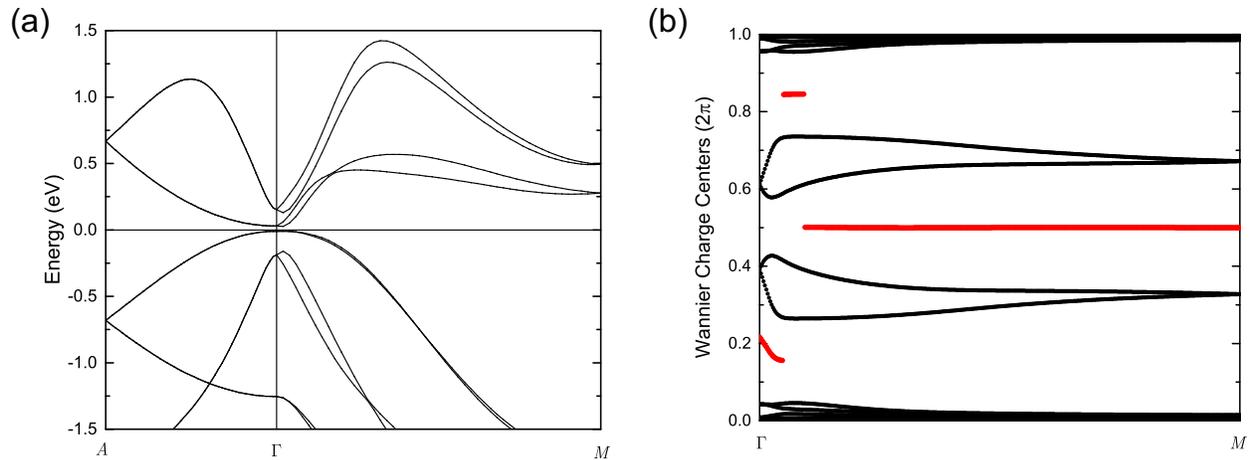}
\caption{(Color online). (a) Band structure around the $\Gamma$ point of LiZnBi in a strained structure with $a=0.97a_0$ and $c=0.98c_0$. (b) WCCs of the strained LiZnBi along $\Gamma$-$M$ at the $\Gamma$-$M$-$K$ plane. The format is the same as that in Fig. \ref{fig:appfig1}.}
\label{fig:appfig2}
\end{figure}

To drive the topological phase transition from 3D Dirac semimetal to convention semiconductor, both $\Gamma_8$ and $\Gamma_7\,(s)$ should be in the conduction bands. Since there is anti-bonding feature in basal plane (along $z$ direction) for $\Gamma_8$ ($\Gamma_7\,(s)$), decreasing $a$ and $c$ at the same time can push up these two energy levels. In fact, when $a=0.97a_0$ and $c=0.98c_0$, we indeed acquire a global gap across the first Brillouin zone and the band structure near the $\Gamma$ point is shown in Fig. \ref{fig:appfig2} (a). In order to calculate the $\mathbb{Z}_2$ index, the WCCs along the $\Gamma$-$M$ line at the $\Gamma$-$M$-$K$ plane are calculated and presented in Fig. \ref{fig:appfig2} (b). The gap center (red dots) cross 6 WCC bands in total as $k$ moves from $\Gamma$ to $M$. Since the crossing number is even, the $\mathbb{Z}_2$ index equals zero and the strained LiZnBi is topologically trivial.

\end{document}